\documentclass[sigconf]{acmart}

\usepackage{amsmath}
\usepackage{bm}
\usepackage{booktabs} %
\usepackage{color}
\usepackage{listings}
\usepackage{setspace}
\usepackage{enumitem}
\usepackage{balance}

\DeclareMathOperator*{\argmin}{argmin}

\copyrightyear{2019} 
\acmYear{2019} 
\setcopyright{rightsretained} 
\acmConference[KDD '19]{The 25th ACM SIGKDD Conference on Knowledge Discovery and Data Mining}{August 4--8, 2019}{Anchorage, AK, USA}
\acmBooktitle{The 25th ACM SIGKDD Conference on Knowledge Discovery and Data Mining (KDD '19), August 4--8, 2019, Anchorage, AK, USA}
\acmDOI{10.1145/3292500.3330677}
\acmISBN{978-1-4503-6201-6/19/08}

\begin{document}
\title{TF-Ranking: Scalable TensorFlow Library for Learning-to-Rank}

\author{
 Rama Kumar Pasumarthi, Sebastian Bruch, Xuanhui Wang, Cheng Li, Michael Bendersky, \\Marc Najork, Jan Pfeifer, Nadav Golbandi, Rohan Anil, Stephan Wolf}
\affiliation{\institution{Google}}
\email{{ramakumar,bruch,xuanhui,chgli,bemike,najork,janpf,nadavg,rohananil,stephanw}@google.com}
\renewcommand{\shorttitle}{TensorFlow Ranking}
\renewcommand{\shortauthors}{Pasumarthi et al.}

\begin{abstract}
 Learning-to-Rank deals with maximizing the utility of a list of examples presented to the user, with items of higher relevance being prioritized. It has several practical applications such as large-scale search, recommender systems, document summarization and question answering. While there is widespread support for classification and regression based learning, support for learning-to-rank in deep learning has been limited. We introduce TensorFlow Ranking, the first open source library for solving large-scale ranking problems in a deep learning framework\footnote{The open source library is available at: \url{https://github.com/tensorflow/ranking}.}. It is highly configurable and provides easy-to-use APIs to support different scoring mechanisms, loss functions and evaluation metrics in the learning-to-rank setting. Our library is developed on top of TensorFlow and can thus fully leverage the advantages of this platform. TensorFlow Ranking has been deployed in production systems within Google; it is highly scalable, both in training and in inference, and can be used to learn ranking models over massive amounts of user activity data, which can include heterogeneous dense and sparse features. We empirically demonstrate the effectiveness of our library in learning ranking functions for large-scale search and recommendation applications in Gmail and Google Drive. We also show that ranking models built using our model scale well for distributed training, without significant impact on metrics. The proposed library is available to the open source community, with the hope that it facilitates further academic research and industrial applications in the field of learning-to-rank.
\end{abstract}



\maketitle

\section{Introduction} \label{sec:intro}

With the high potential of deep learning for real-world data-intensive applications, a number of open source packages have emerged in recent years and are under active development, including TensorFlow~\cite{abadi2016tensorflow}, PyTorch~\cite{paszke2017automatic}, Caffe~\cite{jia2014caffe}, and MXNet~\cite{chen2015mxnet}. Supervised learning is one of the main use cases of deep learning packages. %
However, compared with the comprehensive support for classification or regression in open-source deep learning packages, there is a paucity of support for ranking problems.

A ranking problem is defined as a derivation of ordering over a list of items that maximizes the utility of the entire list. It is widely applicable in several domains, such as Information Retrieval and Natural Language Processing. Some important practical applications include web search, recommender systems, machine translation, document summarization, and question answering~\cite{li2011learning}. 

In general, a ranking problem is different from classification or regression tasks. While the goal of classification or regression is to predict a label or a value for each individual item as accurately as possible, the goal of ranking is to optimally sort the entire item list such that, for some notion of relevance, the items of highest relevance are presented first. To be precise, in a ranking problem, we are more concerned with the relative order of the relevance of items than their absolute magnitudes.

That ranking is a fundamentally different problem entails that classification or regression metrics and methodologies do not transfer effectively to the ranking domain. To fill this void, a number of metrics and a class of methodologies that are inspired by the challenges in ranking have been proposed in the literature. For example, widely-utilized metrics such as Normalized Discounted Cumulative Gain (NDCG)~\cite{jarvelin2002cumulated}, Expected Reciprocal Rank (ERR)~\cite{Chapelle+al:2009}, Mean Reciprocal Rank (MRR)~\cite{craswell2009mean}, Mean Average Precision (MAP), and Average Relevance Position (ARP)~\cite{zhu2004recall} are designed to %
emphasize the items that are ranked higher in the list.

Similarly, a class of supervised machine learning techniques that attempt to solve ranking problems---referred to as learning-to-rank~\cite{li2011learning}---has emerged in recent years. Broadly, the goal of learning-to-rank is to learn from labeled data a parameterized function that maps feature vectors to real-valued scores. During inference, this scoring function is used to sort and rank items. 

Most learning-to-rank methods differ primarily in how they define surrogate loss functions over ranked lists of items during training to optimize a non-differentiable ranking metric, and by that measure fall into one of \emph{pointwise}, \emph{pairwise}, or \emph{listwise} classes of algorithms. Pointwise methods~\cite{Fuhr:1989:OPR:65943.65944,Chu:2005:PLG:1102351.1102369,Gey:1994:IPR:188490.188560} approximate ranking to a classification or regression problem and as such attempt to optimize the discrepancy between individual ground-truth labels and the absolute magnitude of relevance scores produced by the learning-to-rank model. On the other hand, \emph{pairwise}~\cite{burges2005learning,Joachims:2002} or \emph{listwise}~\cite{cao2007learning,xia2008listmle,wang2018lambdaloss} methods either model the pairwise preferences or define a loss over entire ranked list. Therefore, pairwise and listwise methods are more closely aligned with the ranking task~\cite{li2011learning}.

There are other factors that distinguish ranking from other machine learning paradigms. As an example, one of the challenges facing learning-to-rank is the inherent biases~\cite{Joachims+al:2005,Yue:2010:BPB:1772690.1772793} that exist in labeled data collected through implicit feedback (e.g., click logs). Recent work on unbiased learning-to-rank~\cite{ai2018unbiased,Wang+al:2018, Joachims:WSDM17} explores ways to counter position bias~\cite{Joachims+al:2005} in training data and produce a consistent and unbiased ranking function. These techniques work well with pairwise or listwise losses, but not with pointwise losses~\cite{Wang+al:2018}.

From the discussion above, it is clear that a library that supports the learning-to-rank problem has its own unique set of requirements and must offer a functionality that is specific to ranking. Indeed, a number of open source packages such as RankLib\footnote{Available at: \url{https://sourceforge.net/p/lemur/wiki/RankLib/}} and LightGBM~\cite{ke2017lightgbm} exist to address the ranking challenge.

Existing learning-to-rank libraries, however, have a number of important drawbacks. First, they were developed for small data sets (thousands of queries) and do not scale well to massive click logs (hundreds of millions of queries) that are common in industrial applications. Second, they have very limited support for sparse features and can only handle categorical features with a small vocabulary. Crucially, extensive feature engineering is required to handle textual features. In contrast, deep learning packages like TensorFlow can effectively handle sparse features through embeddings~\cite{mikolov2013word2vec}. Finally, existing learning-to-rank libraries do not support the recent advances in unbiased learning-to-rank.

To address this gap, we present our experiences in building a scalable, comprehensive, and configurable industry-grade learning-to-rank library in TensorFlow. Our main contributions are:
\begin{itemize}[leftmargin=*]
  \item We propose an open-source library for training large scale learning-to-rank models using deep learning in TensorFlow. 
  \item The library is flexible and highly configurable: it provides an easy-to-use API to support different scoring mechanisms, loss functions, example weights, and evaluation metrics.
   \item The library provides support for unbiased learning-to-rank by incorporating inverse propensity weights in losses and metrics.
  \item We demonstrate the effectiveness of our library by experiments on two large-scale search and recommendation applications, especially when employing listwise losses and sparse textual features.
  \item We demonstrate the robustness of our library in a large-scale distributed training setting. 
\end{itemize}

Our current implementation of the TensorFlow Ranking library is by no means exhaustive. We envision that this library will provide a convenient open platform for hosting and advancing state-of-the-art ranking models based on deep learning techniques, and thus facilitate both academic research as well as industrial applications.

The remainder of this paper is organized as follows. Section~\ref{sec:ltr} formulates the problem of learning to rank and provides an overview of existing approaches. In Section~\ref{sec:overview}, we present an overview of the proposed learning-to-rank platform, and in Section~\ref{sec:components} we present the implementation details of our proposed library. Section~\ref{sec:eval} showcases uses of the library in a production environment and demonstrates experimental results. Finally, Section~\ref{sec:conclusion} concludes this paper.

\section{Learning-to-Rank} \label{sec:ltr}

In this section, we provide a high-level overview of learning-to-rank techniques. We begin by presenting a formal definition of learning-to-rank and setting up notation.

\subsection{Setup}
Let $\mathcal{X}$ denote the universe of items and let $\bm{x} \in \mathcal{X}^n$ represent a list of $n$ items and $x_i \in \bm{x}$ an item in that list. Further denote the universe of all permutations of size $n$ by $\Pi^n$, where $\pi \in \Pi^n$ is a bijection from $[1:n]$ to itself. $\pi$ may be understood as a total ranking of items in a list where $\pi(i)$ yields the rank according to $\pi$ of the $i^{\text{th}}$ item in the list and $\pi^{-1}(r)$ yields the index of the item at rank $r$, and we have that $\pi^{-1}(\pi(i)) = i$. A ranking function $f: \mathcal{X}^n \rightarrow \Pi^n$ is a function that, given a list of $n$ items, produces a permutation or a ranking of that list.

The goal of learning-to-rank, in broad terms, is to learn a ranking function $f$ from training data such that items as ordered by $f$ yield maximal utility. Let us parse this statement and discuss each component in more depth.

\subsection{Training Data}
We begin with a description of training data. Learning-to-rank, which is an instance of supervised learning, assumes the existence of a ground-truth permutation $\pi^\ast$ for a given list of items $\bm{x}$.

In most real-world settings, ground-truth is provided in its more general form: a \emph{set} of permutations or in other words a \emph{partial} ranking. In a partial ranking $X_1 \succ X_2 \succ ... \succ X_k$, where $X_i$s are $k$ disjoint subsets of elements of $\bm{x} \in \mathcal{X}^n$ ($k \leq n$), items in $X_i$ are preferred over those in $X_{j>i}$, but within each $X_i$ items may be permuted freely. When $k=n$ partial ranking reduces to a total ranking.

As a concrete example, consider the task of \emph{ad hoc} retrieval where given a (textual) query the ranking algorithm retrieves a relevant list of documents from a large corpus. When constructing a training dataset, one may recruit human experts to examine an often very small subset of candidate documents for a given query and grade the documents' relevance with respect to that query on some scale (e.g., 0 for "not examined" or "not relevant" to 5 for "highly relevant"). A relevance grade for document $x_i$ is considered its "label" $y_i$. Similarly, in training datasets that are constructed by implicit user feedback such as click logs, documents are either relevant and clicked ($y=1$) or not ($y=0$). In either case, a list of labels $\bm{y}$ induces a partial ranking of documents.

In order to simplify notation throughout the remainder of this paper and without loss of generality, we assume the existence of $\pi^\ast$, a correct total ranking. $\pi^\ast$ can be understood as an ranking induced by a list of labels $\bm{y}\in\mathbb{R}^n$ for $\bm{x}\in\mathcal{X}^n$. As such, our training data set of $m$ items can be defined as $S^m = \{(\bm{x}, \pi^\ast)\, |\, \bm{x}\in\mathcal{X}^n, \pi^\ast\in\Pi^n\}$ or equivalently $S^m = \{(\bm{x}, \bm{y}) \in \mathcal{X}^n\times\mathbb{R}^n\}$.

Returning to the case of \emph{ad hoc} retrieval, it is worth noting that each item $x_i \in \bm{x}$ is in fact a pair of query and document $(q, d_i)$: It is generally the case that the pair $(q, d_i)$ is transformed to a feature vector $x_i$.

\subsection{Scoring Function}
\label{sec:scoring-function}
Directly finding a permutation $\pi$ is difficult, as the space of all possible permutations is exponentially large. In practice, a score-and-sort approach is used instead. Let $h : \mathcal{X}^n \rightarrow \mathbb{R}^n$ be a scoring function that maps a list of items $\bm{x}$ to a list of scores $\hat{\bm{y}}$. Let $h(.)|_k$ denotes the $k^{\text{th}}$ dimension of $h(.)$. As discussed earlier, $h$ induces a permutation $\pi$ such that $h(\bm{x})|_{\pi^{-1}(r)}$ is monotonically decreasing for increasing ranks $r$. 

In its simplest form, the scoring function is univariate and can be decomposed into a per-item scoring function as shown in Equation~\ref{eq:pointwise_scoring}, where $g: x \rightarrow \mathbb{R}$ maps a feature vector to a real-valued score.
\begin{equation}
\centering
 h(\bm{x}) = [g(x_i),\, \forall\, 1\leq i \leq n]
         = [g(x_1),\, g(x_2),\, ...,\, g(x_n)].
 \label{eq:pointwise_scoring}
\end{equation}

The scoring function $h$ is typically parameterized by a set of parameters $\theta$ and can be written as $h(.; \theta)$. Many parameterization options have been studied in the learning-to-rank literature including linear functions~\cite{joachims2006training}, boosted weak learners~\cite{Jun+Hang:2007}, gradient-boosted trees~\cite{friedman2001greedy, burges2010ranknet}, support vector machines~\cite{joachims2006training, Joachims:WSDM17}, and neural networks~\cite{burges2005learning}. Our library offers deep neural networks as the basis to construct a scoring function. This framework facilitates more sophisticated scoring functions such as multivariate functions~\cite{ai2018groupwise}, where the scores of a group of items are computed jointly. Through a flexible API, the library also enables development and integration of arbitrary scoring functions into a ranking model.

\subsection{Utility and Ranking Metrics}
\label{sec:metrics}
We now turn to the notion of utility. As noted in Section~\ref{sec:intro}, the utility of an ordered list of items is often measured by a number of standard ranking-specific metrics. What makes ranking metrics unique and suitable for this task is that, in ranking, it is often desirable to have fewer errors at higher ranked positions; this principle is reflected in many ranking metrics:

\begin{equation}
\mathit{RR}(\pi, \bm{y}) = \frac{1}{\min_{j}\{y_{\pi^{-1}(j)} > 0\}},  
\end{equation}
\begin{equation}
  \mathit{RP}(\pi, \bm{y}) = \frac{\sum_{j=1}^n y_j \pi(j)}{\sum_{j=1}^n y_j},
\end{equation}

\begin{equation}
  \mathit{DCG}(\pi, \bm{y}) =\sum_{j=1}^n \frac{2^{y_j}-1}{\log_2(1 + \pi(j))}, 
\end{equation}
\begin{equation}
  \mathit{NDCG}(\pi, \bm{y}) = \frac{\mathit{DCG(\pi, \bm{y})}}{\mathit{DCG}(\pi^\ast, \bm{y})},
\end{equation}
where $y_i \in \bm{y}$ are ground-truth labels that induce $\pi^\ast$, and $\pi(i)$ is the rank of the $i^{\text{th}}$ item in $\bm{x}$. $\mathit{RR}$ is the reciprocal rank of the first relevant item. $\mathit{RP}$ is the positions of items weighted by their relevance values~\cite{zhu2004recall}. DCG is the Discounted Cumulative Gain~\cite{jarvelin2002cumulated}, and NDCG is DCG normalized by the maximum DCG obtained from the ideal ranked list $\pi^\ast$.

Note that, given $m$ evaluation samples ${(\pi_k, \bm{y}_k), 1 \leq k \leq m}$, the mean of the above metrics is calculated and reported instead. For example, the mean reciprocal rank (MRR) is defined as:
$$MRR = \frac{1}{m} \displaystyle\sum_{k=1}^N RR(\pi_k, \bm{y}_k).$$
Our library supports commonly used ranking metrics and enables easy development and addition of arbitrary metrics.

\subsection{Loss Functions}
\label{sec:losses}
Learning-to-rank seeks to maximize a utility or equivalently minimize a cost or loss function. Assuming there exists a loss function $\ell(.)$, the objective of learning-to-rank is to find a ranking function $f^\ast$ that minimizes the empirical loss over training samples:

\begin{equation}
    \displaystyle{f^\ast = \argmin_{f: \mathcal{X}^n \rightarrow \Pi^n} \dfrac{1}{m} \sum_{(\bm{x}, \pi^\ast) \in S^m} {\ell(\pi^\ast, f(\bm{x}))} }.
\end{equation}

Replacing $f$ with a scoring function $h$ as is often the case yields the following optimization problem:

\begin{equation}
    \displaystyle{h^\ast = \argmin_{h: \mathcal{X}^n \rightarrow \mathbb{R}^n} \dfrac{1}{m} \sum_{(\bm{x}, \bm{y}) \in S^m} {\hat{\ell}(\bm{y}, h(\bm{x}))} },
\end{equation}
where $\hat{\ell(.)}$ is a loss function equivalent to $\ell(.)$ that acts on scores instead of permutations induced by scores.

This setup naturally prefers loss functions that are differentiable. Most ranking metrics, however, are either discontinuous or flat everywhere due to the use of the sort operation and as such cannot be directly optimized by learning-to-rank methods. With a few notable exceptions~\cite{metzler2005directMaximization,Jun+Hang:2007}, most learning-to-rank approaches therefore define and optimize a differentiable surrogate loss instead. They do so by, among other techniques, creating a smooth variant of popular ranking metrics~\cite{Qin:2010:GAF:1842549.1842572,Taylor+al:2008}; deriving tight upper-bounds on ranking metrics~\cite{wang2018lambdaloss}; bypassing the requirement that a loss function be defined altogether~\cite{burges2010ranknet}; or, otherwise designing losses that are loosely related to ranking metrics~\cite{xia2008listmle,cao2007learning}.

Our library supports a number of surrogate loss functions. As an example of a \textbf{pointwise} loss, the sigmoid cross-entropy for binary relevance labels $y_j \in \{0, 1\}$ is computed as follows:

\begin{equation}
\hat{\ell}(\bm{y}, \hat{\bm{y}}) = - \displaystyle\sum_{j=1}^n  y_j \log(p_j) + (1-y_j) \log(1-p_j)
\end{equation}
where $p_j = \frac{1}{1 + \exp(-\hat{y}_j)}$, $\hat{\bm{y}} \triangleq h(\bm{x})$ are scores computed by the scoring function $h$. As an example of a \textbf{pairwise} loss in our library, the pairwise logistic loss is defined as: 

\begin{equation}
\hat{\ell}(\bm{y}, \hat{\bm{y}}) = \displaystyle\sum_{j=1}^n \displaystyle\sum_{k=1}^n  \mathbb{I}(y_j > y_k) \log(1 + \exp(\hat{y}_k - \hat{y}_j)))
\label{eq:logistic_loss}
\end{equation}
where $\mathbb{I}(\cdot)$ is the indicator function. Finally, as an example of a \textbf{listwise} loss~\cite{NIPS2009_3708}, our library provides the implementation of Softmax Cross-Entropy, ListNet~\cite{cao2007learning}, and ListMLE~\cite{xia2008listmle} among others. For example, the Softmax Cross-Entropy loss is defined as follows:

\begin{equation}
\hat{\ell}(\bm{y}, \hat{\bm{y}}) = - \displaystyle\sum_{j=1}^n y_j \log(\frac{\exp(\hat{y}_j)}{\sum_{j=1}^n \exp(\hat{y}_j)})
\end{equation}

\subsection{Item Weighting}
\label{sec:item_weights}
Finally, we conclude this section by a note on bias in learning-to-rank. As discussed in Section~\ref{sec:intro}, a number of studies have shown that click logs exhibit various biases including position bias~\cite{Joachims+al:2005}. In short, users are less likely to examine and click items at larger rank positions. Ignoring this bias when training a learning-to-rank model or when evaluating ranked lists may lead to a model with less generalization capacity and inaccurate quality measurements.

Unbiased learning-to-rank~\cite{Joachims:WSDM17, Wang+al:2018} looks at handling such biases in relevance labels. One proposed method~\cite{Wang+al:2016, Wang+al:2018} is to compute Inverse Propensity Weights (IPW) for each position in the ranked list. By incorporating these scores in the training process (usually by way of re-weighting items during loss computation), one may produce a better ranking function. Similarly, IPW-weighted variants of evaluation metrics attempt to counter such biases during evaluation. Our library supports the incorporation of such weights into the training and evaluation processes.

\section{Platform Overview}
\label{sec:overview}

Popular use-cases of learning-to-rank, such as search or recommender systems~\cite{tata2017quick}, have several challenges. 
Models are generally trained over vast amounts of user data, so efficiency and scalability are critical. 
Features may be comprised of dense, categorical, or sparse types, and are often missing for some data points. These applications also require fast inference for real-time serving. 

It is against the backdrop of these challenges that we believe, at a high level, neural networks as a class of machine learning models, and TensorFlow~\cite{abadi2016tensorflow} as a machine learning framework are suitable for practical learning-to-rank problems. 

Take efficiency and scalability as an example. The availability of vasts amounts of training data, along with an increase in computational power and thereby in our ability to train deeper neural networks with millions of parameters in a scalable manner have led to rapid adoption of neural networks for a variety of applications. In fact, deep neural networks have gained immense popularity, with applications in Natural Language Processing, Computer Vision, Speech Signal Processing, and many other areas~\cite{goodfellow2016deep}. Recently, neural networks have been shown to be effective in applications of Information Retrieval as well~\cite{mitra2017neural}.

Neural networks can also process heterogeneous features more naturally. Evidence~\cite{bengio2013representation} suggests that neural networks learn effective representations of categorical and sparse features. These include words or characters in Natural Language Processing~\cite{mikolov2013word2vec}, phonemes in Speech Processing~\cite{silfverberg2018sound}, or raw pixels in Computer Vision~\cite{lecun1995convolutional}. Such "representation learning" is usually achieved by way of converting raw, unstructured data into a dense real valued vector, which, in turn, is treated as a vector of implicit features for subsequent neural network layers.
For example, discrete elements such as words or characters from a finite vocabulary are transformed using an embedding matrix---also referred to as embedding layer or embeddings~\cite{bengio2013representation}---which maps every element of the vocabulary to a learned, dense representation. This particular representation learning is useful for learning-to-rank over documents, web pages or other textual data.

Given these properties, neural networks are excellent candidates for modeling a score-and-sort approach to learning-to-rank. In particular, the scoring function $h(.)$ in Equation~\ref{eq:pointwise_scoring} can be parameterized by a neural network. In such a setup, feature representations can be jointly learned with the parameters of $h(.;\, \theta)$ while minimizing the objective $\hat{\ell}$ averaged over the training data. This is the general setup we adopt and implement in the TensorFlow Ranking library.

TensorFlow Ranking is built on top of TensorFlow, a popular open-source library for large scale training, evaluation and serving of machine learning and deep learning models.
TensorFlow supports high performance tensor (multi-dimensional vector) manipulation and computation via what is referred to as "computational graphs." A computational graph expresses the logic of a sequence of tensor manipulations, with each node in the graph corresponding to a single tensor operation (\textit{op}). For example, a node in the computation graph may multiply an "input" tensor with a "weight" tensor, a subsequent node may add a "bias" tensor to that product, and a final node may pass the resultant tensor through the Sigmoid function.

The concept of a computation graph with tensor operations simplifies the implementation of the \textit{backpropagation}~\cite{chauvin2013backpropagation} algorithm for training neural networks. In a forward pass, the values at each node are computed by composing a sequence of tensor operations, and in a backward pass, gradients are accumulated in the the reverse fashion. 
TensorFlow enables such propagation of gradients through automatic differentiation~\cite{abadi2016tensorflow}: each operation in the computation graph is equipped with a gradient expression with respect to its input tensors. In this way, the gradient of a complex composition of tensor operations can be automatically inferred during a backward pass through the computation graph. This allows for composition of a large number of operations to construct deeper networks.

Another computationally attractive property of the TensorFlow framework is its support, via TensorFlow Estimator \cite{cheng2017tensorflow}, of distributed training of neural networks. TensorFlow Estimator is an abstract library that takes a high-level training, evaluation, or prediction logic and hides the execution logic from developers. An \texttt{Estimator} object encapsulates two major abstract components that can be further customized: (1) \texttt{input\_fn}, which reads in data from a persistent storage and creates tensors for features and labels, and (2) \texttt{model\_fn} which processes input features and labels, and depending on the mode (\texttt{TRAIN}, \texttt{EVAL}, \texttt{PREDICT}), returns a loss value, evaluation metrics, or predictions.
The computation graph expressed within the \texttt{model\_fn} may depend on the mode. This is particularly useful for learning-to-rank because the model may need entire lists of items during training (to compute a listwise loss, for example), but during serving, it may score each item independently.
The model function itself can be expressed as a combination of a \texttt{logits} builder and a \texttt{Head} abstraction, where the logits builder generates the values in the forward computation of the graph, and the \texttt{Head} object defines loss objectives and associated metrics.

These abstractions, along with modular design, the ability to distribute the training of a neural network and to serve on a variety of high-performance hardware platforms (CPU/GPU/TPUs) are what make the TensorFlow ecosystem a suitable platform for a neural network-based learning-to-rank library. In the next section, we discuss how the design principles in TensorFlow and the \texttt{Estimator} workflow inspire the design of the TensorFlow Ranking library.

\begin{figure*}
\centering
\fbox{
\includegraphics[width=0.8\linewidth,keepaspectratio]{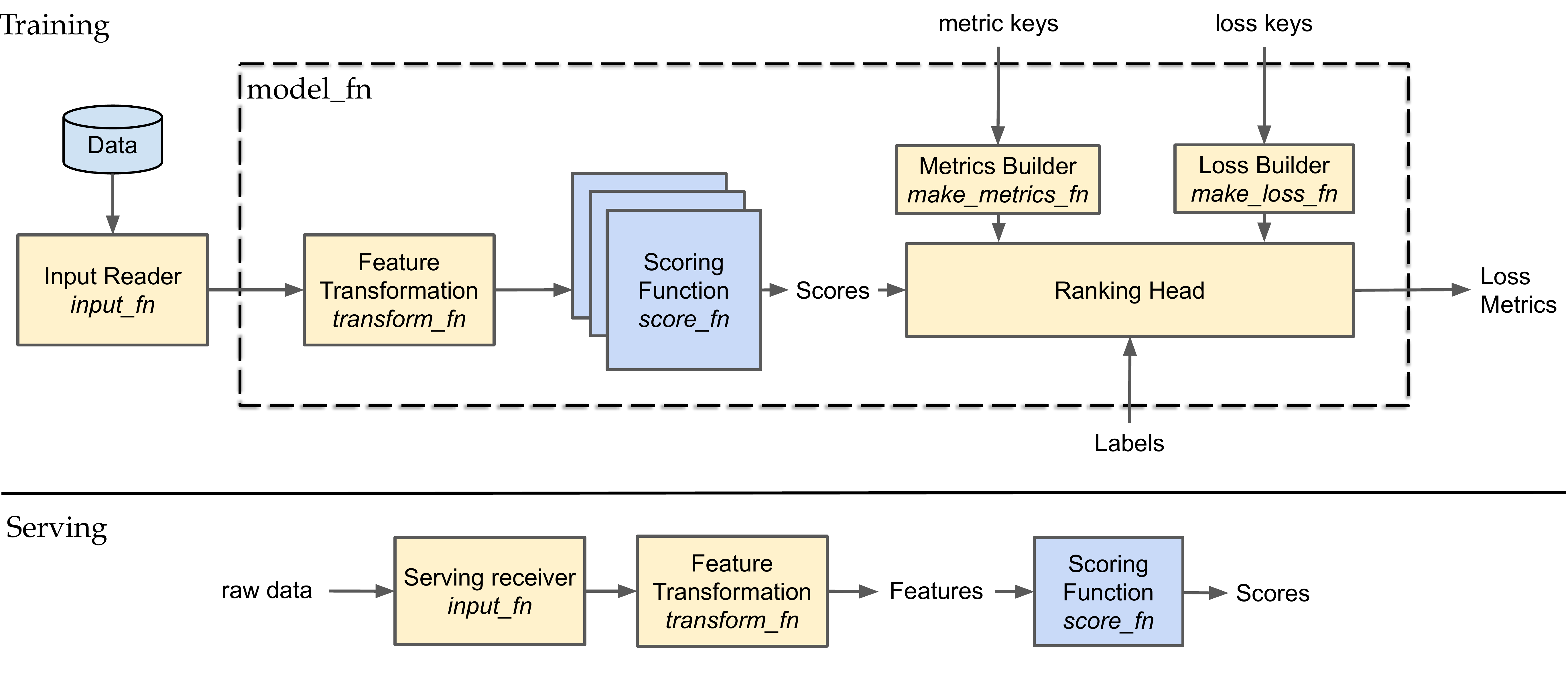}
}
\caption{TensorFlow Ranking Architecture}
\label{fig:arch}
\end{figure*}

\section{Components}
\label{sec:components}
Motivated by design patterns in TensorFlow and the \texttt{Estimator} framework, TensorFlow Ranking constructs computational sub-graphs via callbacks for various aspects of a learning-to-rank model such as scoring function $h(.)$, losses $\hat{\ell}(.)$, and evaluation metrics (see Section~\ref{sec:ltr}).
These subgraphs are combined together in the \texttt{Estimator} framework, via a custom callback to create a \texttt{model\_fn}. 
The \texttt{model\_fn} itself can be decomposed into a \texttt{logits} builder, which represents the output of the neural network layers, and a \texttt{Head} abstraction, which encapsulates loss and metrics.
This decomposition is particularly suitable to learning-to-rank problems in the score-and-sort approach. The \texttt{logits} builder corresponds to the scoring function, defined in Section \ref{sec:scoring-function}, and the \texttt{Head} corresponds to losses and metrics. 

It is important to note that this decomposition also provides modularity and the ability to switch between various combinations of scoring functions and ranking heads. For example, as we show in the code examples below, switching between a pointwise or a listwise loss, or incorporating embedding for sparse features into the model can be expressed in a single line code change. We found this modularity to be valuable in practical applications, where rapid experimentation is often required.

The overall architecture of TensorFlow Ranking library is illustrated in Figure \ref{fig:arch}. 
The key components of the library are: (1) data reader, (2) transform function, (3) scoring function, (4) ranking loss functions, (5) evaluation metrics, (6) ranking head, and (7) a \texttt{model\_fn} builder. In the remainder of this section, we will describe each component in detail.

\subsection{Reading data using \texttt{input\_fn}}
The \texttt{input\_fn}, as shown in Figure~\ref{fig:arch}, reads in data either from a persistent storage or data generated by another process, to produce dense and sparse tensors of the appropriate type, along with the labels. The library provides support for several popular data formats (\texttt{LIBSVM, tf.SequenceExample}). Furthermore, it allows developers to define customized data readers.

Consider the case when the user would like to construct a custom \texttt{input\_fn}. The user defines a parser for a batch of serialized datapoints that returns a feature dictionary containing 2-D and 3-D tensors for "context" and "per-item" features respectively. This user-defined parser function is passed to a dataset builder, which generates features and labels. Per-item features are those features that are particular to each example or item in the list of items. Context features, on the other hand, are independent of the items in the list. In the context of document search, features extracted from documents are per-item features and other document-independent features (such as query/session/user features) are considered context features. We represent per-item features by a 3-D tensor (where dimensions correspond to queries, items, and feature values) and context features by a 2-D tensor (where dimensions correspond to queries and feature values). The following code snippet shows how one may construct a custom \texttt{input\_fn}.

\noindent
\begin{lstlisting}[language=Python,breaklines=true,frame=single]
def _parse_single_datapoint(serialized):
  """User defined logic to parse a batch of serialized datapoints to a dictionary of 2-D and 3-D tensors."""
  return features

def input_fn(file_pattern):
  """Generate features and labels from input data."""
  dataset = tfr.data.build_ranking_dataset_with_parsing_fn(
  file_pattern=file_pattern, 
  parsing_fn=_parse_single_datapoint, ...)
  features = dataset.make_one_shot_iterator().get_next()
  label = tf.squeeze(features.pop(_LABEL_FEATURE), axis=2)
  return features, label
\end{lstlisting}

\subsection{Feature Transformation with \texttt{transform\_fn}}
As discussed in Section~\ref{sec:overview}, sparse features such as words or n-grams can be transformed to dense features using (learned) embeddings. More generally, any raw feature may require some form of transformation. Such transformations may be implemented in \texttt{transform\_fn}, a function that is applied to the output of \texttt{input\_fn}. The library provides standard functions to transform sparse features to dense features based on feature definitions. A feature definition in TensorFlow is a user-defined \texttt{tf.FeatureColumn}~\cite{cheng2017tensorflow}, an expression of the type and attributes of features. Given feature definitions, the transform function produces dense 2-D or 3-D tensors for context and per-item features respectively. The following code snippet demonstrates an implementation of the \texttt{transform\_fn}. Context and per-item feature definitions are passed to the \texttt{encode\_listwise\_features} function, which in turn returns dense tensors for each of context and per-item features.

\noindent
\begin{minipage}{\linewidth}
\begin{lstlisting}[language=Python,breaklines=true,frame=single]
def make_transform_fn():
  def _transform_fn(features, mode):
    context_feature_columns = {"unigrams": embedding_column(
      categorical_column("unigrams", dimension=10, vocab_list=_VOCAB_LIST))}
    example_feature_columns = {"utility": numeric_column("utility", 
        shape=(1,), default_value=0.0, dtype=float32)}
    # 2-D context tensors and 3-D per-item tensors.
    context_features, example_features = tfr.feature.encode_listwise_features(
        features, 
        input_size=2,
        context_feature_columns=context_feature_columns,
        example_feature_columns=example_feature_columns)
    return context_features, example_features
  return _transform_fn
\end{lstlisting}
\end{minipage}
\vspace{-5pt}

\subsection{Feature Interactions using \texttt{scoring\_fn}}
The library allows users to build arbitrary scoring functions defined in Section~\ref{sec:scoring-function}. A scoring function takes batched tensors in the form of 2-D context features and 3-D per-item features, and returns a score for a single item or a group of items.
The scoring function is supplied to the model builder via a callback. The model uses the scoring function internally to generate scores during training and inference, as shown in Figure~\ref{fig:arch}. The signature of the scoring function is shown below, where \texttt{mode}, \texttt{params}, and \texttt{config} are input arguments for \texttt{model\_fn} that supply model hyperparameters and configuration for distributed training~\cite{cheng2017tensorflow}. The code snippet below, constructs a 3-layer feedforward neural network with ReLUs~\cite{nair2010rectified}.

\noindent
\begin{minipage}{\linewidth}
\begin{lstlisting}[language=Python,breaklines=true,frame=single]
def make_score_fn():
  def _score_fn(context_features, group_features, mode, params, config):
    """Scoring network for feature interactions."""
    net = concat(layers.flatten(group_features.values()), layers.flatten(context_features.values()))
    for i in range(3):
      net = layers.dense(net, units=128, activation="relu")
    logits = layers.dense(cur_layer, units=1)
    return logits
  return _score_fn
\end{lstlisting}
\end{minipage}
\vspace{-5pt}
\vspace{-5pt}

\subsection{Ranking Losses}
\label{sec:losses_api}
Here, we describe the APIs for building loss functions defined in Section \ref{sec:losses}. 
Losses in TensorFlow are functions that take in inputs, labels and a weight, and return a weighted loss value. 
The library has a pre-defined set of pointwise, pairwise and listwise ranking losses. 
The loss key is an enum over supported loss functions. 
These losses are exposed using the factory function \texttt{tfr.losses.make\_loss\_fn} that takes a loss key (name) and a weights tensor, and returns a loss function compatible with \texttt{Estimator}.
The code snippet below shows how to built the loss function for softmax cross-entropy loss.

\noindent
\begin{minipage}{\linewidth}
\begin{lstlisting}[language=Python,breaklines=true,frame=single]
# Define loss key(s).
loss_key = tfr.losses.RankingLossKey.SOFTMAX_LOSS
# Build the loss function.
loss_fn = tfr.losses.make_loss_fn(loss_key, ...)
# Generating loss value from the loss function.
loss_scalar = loss_fn(scores, labels, ...)
\end{lstlisting}
\end{minipage}
\vspace{-5pt}

\subsection{Ranking Metrics}
\label{sec:metrics_api}
The library provides an API to compute most common ranking metrics defined in Section \ref{sec:metrics}. %
Similar to loss functions, a metric can be instantiated using a factory function that takes a metric key and a weights tensor, and returns a metric function compatible with \texttt{Estimator}. The metric function itself takes in predictions, labels, and weights to compute a scalar measure. The metric key is an enum over supported metric functions, described in Section \ref{sec:metrics}. 
During evaluation, the library supports computing both weighted and unweighted metrics, for example weights defined in Section~\ref{sec:item_weights}, which facilitates evaluation in the context of unbiased learning-to-rank. The code snippet below shows how to build metric function. 

\noindent
\begin{minipage}{\linewidth}
\begin{lstlisting}[language=Python,breaklines=true,frame=single]
def eval_metric_fns():
  """Returns a dict from name to metric functions."""
  metric_fns = {
      "metric/ndcg@5": tfr.metrics.make_ranking_metric_fn(
          tfr.metrics.RankingMetricKey.NDCG, topn=5)
  }
  return metric_fns
\end{lstlisting}
\end{minipage}

\vspace{-5pt}
\subsection{Ranking Head} 
In the \texttt{Estimator} workflow, the \texttt{Head} API is an abstraction that encapsulates losses and metrics: Given a pair of \texttt{Estimator}-compatible loss function and metric function along with scores from a neural network, \texttt{Head} computes the values of the loss and metric and produces model predictions as output.
The library provides a \texttt{Ranking} \texttt{Head}: a \texttt{Head} object with built-in support for ranking losses built in Section~\ref{sec:losses_api} and ranking metrics of Section~\ref{sec:metrics_api}. The signature of \texttt{Ranking} \texttt{Head} is shown below.

\noindent
\begin{minipage}{\linewidth}
\begin{lstlisting}[language=Python,breaklines=true,frame=single]
def _train_op_fn(loss):
  """Defines train op used in ranking head."""
  return tf.contrib.layers.optimize_loss(
      loss=loss,
      global_step=tf.train.get_global_step(),
      learning_rate=hparams.learning_rate,
      optimizer="Adagrad")
      
ranking_head = tfr.head.create_ranking_head(
    loss_fn=tfr.losses.make_loss_fn(_LOSS),
    eval_metric_fns=eval_metric_fns(),
    train_op_fn=_train_op_fn)
\end{lstlisting}
\end{minipage}
\vspace{-5pt}

\subsection{Model Builder} 
A model builder, \texttt{model\_fn}, is what puts all the different pieces together: scoring function, transform function, losses and metrics via ranking head. 
Recall that the \texttt{model\_fn} returns operations related to predictions, metrics, and loss optimization. The output of \texttt{model\_fn} and the graph constructed depends on the mode \texttt{TRAIN}, \texttt{EVAL}, or \texttt{PREDICT}. These are all handled internally by the library through \texttt{make\_groupwise\_ranking\_fn}.
The signature of a model builder, along with the overall flow to build a ranking \texttt{Estimator} is shown in Figure~\ref{fig:arch}.
The components of the ranking library can be used to construct a ranking model in several ways. The inbuilt model builder, \texttt{make\_groupwise\_ranking\_fn}, provides a ranking model with multivariate scoring function and configurable losses and metrics. The user can also define a custom model builder which can use components from the library, such as losses or metrics. 

\noindent
\begin{minipage}{\linewidth}
\begin{lstlisting}[language=Python,breaklines=true,frame=single]
input_fn = tfr.data.read_batched_sequence_example_dataset(file_pattern,...)
ranking_estimator = estimator.Estimator(
      model_fn=tfr.model.make_groupwise_ranking_fn(
          group_score_fn=make_score_fn(),
          group_size=group_size,
          transform_fn=make_transform_fn(),
          ranking_head=ranking_head),
      params=hparams)
# Training loop.
for _ in range(num_train_steps):
  ranking_estimator.train(input_fn(TRAIN_FILES),...)
# Evaluation.
ranking_estimator.evaluate(input_fn(EVAL_FILES),...)
\end{lstlisting}
\end{minipage}

\vspace{5pt}
Ranking models have a crucial training-serving discrepancy. During training the model receives a list of items, but during serving, it could potentially receive independent items which are generated by a separate retrieval algorithm. The ranking model builder handles this by generating a graph compatible with serving requirements and exporting it as a \texttt{SavedModel}~\cite{olston2017tensorflow}: a language agnostic object that can be loaded by the serving code, which can be in any low-level language such as C++ or Java.

\vspace{-3pt}
\section{Use Cases} \label{sec:eval}

Tensorflow Ranking is already deployed in several production systems at Google. In this section, we demonstrate the effectiveness of our library for two real-world ranking scenarios: \emph{Gmail search}~\cite{Wang+al:2016,Zamani+al:2017} and \emph{document recommendation in Google Drive}~\cite{tata2017quick}. In both cases the model is trained on large quantities of click data that is beyond the capabilities of existing open source learning-to-rank packages, e.g., RankLib. In addition, in the Gmail setting, our model contains sparse textual features that cannot be naturally handled by the existing learning-to-rank packages. 

\vspace{-3pt}
\subsection{Gmail Search}
In one set of experiments, we evaluate several ranking models trained on search logs from Gmail. In this service, when a user types a query into the search box, five results are shown and user clicks (if any) are recorded and later used as relevance labels. To preserve user privacy, we remove personal information and anonymize data using $k$-anonymization. We obtain a set of features that consists of both dense and sparse features. Sparse features include word- and character-level n-grams derived from queries and email subjects. The vocabulary of n-grams is pruned to retain only n-grams that occur across more than $k$ users. This is done to preserve user privacy, as well as to promote a common vocabulary for learning a shared representations across users. %
In total, we collect about 250M queries and isolate 10\% of those to construct an evaluation set. Losses and metrics are weighted by Inverse Propensity Weighting~\cite{Wang+al:2016} computed to counter position bias.

\begin{table}
\caption{Model performance with various loss functions. $\Delta M$ denotes \% improvement in metric $M$ over the \emph{Sigmoid Cross Entropy} baseline. 
Best performance per column is in bold.}
\vspace{-10pt}
\label{tab:gmail}
\centering
\begin{tabular}{@{}llll@{}}
\toprule
(a) Gmail Search                          & $\Delta$MRR      &  $\Delta$ARP     & $\Delta$NDCG    \\ \midrule
Sigmoid Cross Entropy (Pointwise)     & --      & --  & --    \\ 
Logistic Loss (Pairwise)    & +1.52       & +1.64  & +1.00    \\
Softmax Cross Entropy (Listwise)     & {\bf +1.80}    & {\bf +1.88}  & {\bf +1.57}   \\  \midrule \midrule
(b) Quick Access                          & $\Delta$MRR      &  $\Delta$ARP     & $\Delta$NDCG    \\ \midrule
Sigmoid Cross Entropy (Pointwise)     & --     & --  & --   \\ 
Logistic Loss (Pairwise)    & +0.70       & +1.86  & +0.35   \\ 
Softmax Cross Entropy (Listwise)     & {\bf +1.08}   & {\bf +1.88}  & {\bf +1.05}  \\ \bottomrule

\end{tabular}
\end{table}

\vspace{-3pt}
\subsection{Document Recommendation in Drive}
Quick Access in Google Drive~\cite{tata2017quick} is a zero-state recommendation engine that surfaces documents currently relevant to the user when she visits the Drive home screen. 
We evaluate several ranking models trained on user click data over these recommended results. The set of features consists of mostly dense features, as described in Tata et al.~\cite{tata2017quick}. %
In total we collected about 30M instances and set aside 10\% of the set for evaluation.

\subsection{Model Effectiveness}
\subsubsection{Setup and Evaluation}
We consider a simple 3-layer feed-forward neural network with ReLU \cite{nair2010rectified} non-linear activation units and dropout regularization \cite{srivastava2014dropout}. We train models using pointwise, pairwise, and listwise losses defined in Section \ref{sec:losses},  and use \texttt{Adagrad} \cite{duchi2011adaptive} to optimize the objective. We set the learning rate to 0.1 for Quick Access, and 0.3 for Gmail Search.

The models are evaluated using the metrics defined in Section \ref{sec:metrics}.
Due to the proprietary nature of the models, we only report relative improvements with respect to a given baseline. Due to the large size of the evaluation datasets, all the reported improvements are statistically significant.

\subsubsection{Effect of listwise losses}
Table~\ref{tab:gmail} summarizes the impact of different loss functions as measured by various ranking metrics for Gmail and Drive experiments respectively. We observe that a listwise loss performs better than a pairwise loss, which is in turn better than a pointwise loss. This observation confirms the importance of listwise losses for ranking problems over pointwise and pairwise losses, both in search and recommendation settings.

\subsubsection{Incorporating sparse features}

Unlike other approaches to learning to rank, like linear models, SVMs, or GBDTs, neural networks can effectively incorporate sparse features like query or document text. Neural networks handle sparse features by using embedding layers, which map each sparse value to a dense representation. These embedding matrices can be jointly trained along with the neural network parameters, allowing us to learn effective dense representations for sparse features in the context of a given task (e.g., search or recommendation). As prior work shows, these representations can substantially improve model effectiveness on large-scale collections \cite{Xiong+al:2017}.

To demonstrate this point, Table ~\ref{tab:sparse_features} reports the relative improvements from using sparse features in addition to dense features on Gmail Search. We use an embedding layer of size 20 for each sparse feature. We note that adding sparse features significantly boosts ranking quality across metrics and loss functions. This confirms the importance of sparse textual features in large-scale applications, and the effectiveness of TensorFlow Ranking in employing these features.

\begin{figure}[t]
\centering
\includegraphics[width=0.8\linewidth,keepaspectratio]{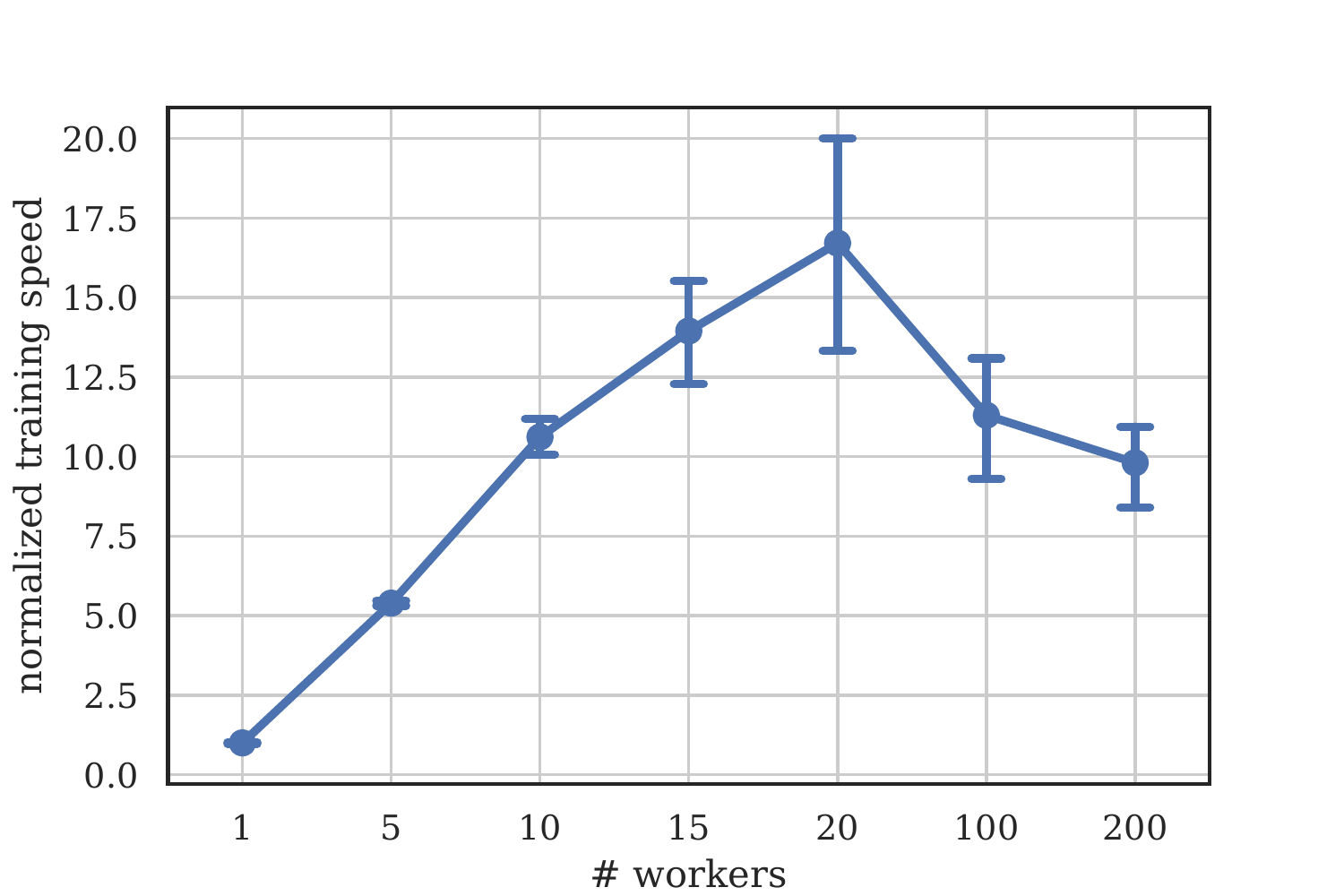}
\caption{Normalized training speed for Gmail Search as a function of number of workers.}
\label{fig:scale-speed}
\vspace{-8pt}
\end{figure}

\begin{table}[b]
\vspace{-3mm}
\caption{Model performance with dense and sparse textual features. $\Delta M$ denotes \% improvement in metric $M$ over the corresponding baseline, when only dense features are used. }
\vspace{-10pt}
\label{tab:sparse_features}
\centering
\begin{tabular}{@{}llll@{}}
\toprule
                                      & $\Delta$MRR     &  $\Delta$ARP     & $\Delta$NDCG    \\ \midrule
Sigmoid Cross Entropy (Pointwise)      & +6.06        & +6.87          & +3.92         \\ 
Logistic Loss (Pairwise)               & +5.40        & +6.25          & +3.51         \\ 
Softmax Cross Entropy (Listwise)       & +5.69        & +6.25          & +3.70         \\ \bottomrule
\end{tabular}
\end{table}

\subsection{Distributed Training}
Distributed training is crucial for applications of learning-to-rank to large scale datasets. Models built using the \texttt{Estimator} class allow for easy switching between local and distributed training, without changing the core functionality of the model. This is achieved via the \texttt{Experiment} class~\cite{cheng2017tensorflow} given a distribution strategy. 

Distributed training using TensorFlow consists of a number of worker tasks and parameter servers. We investigate the scalability of the model built for Gmail Search, the larger of our two datasets. We examine the effect of increasing the number of workers on the training speed. 
For the distributed strategy, we use between-graph replication, where each worker replicates the graph on a different subset of the data, and asynchronous training, where the gradients from different workers are asynchronously collected to update the gradients for the model.

For each of the following experiments, the number of workers ranges between 1 and 200, while the number of training epochs is fixed at 20 million. For robustness, each configuration is run 5 times, and the $95\%$ confidence intervals are plotted.

\subsubsection{Effect of scaling on training speed}
We look at the impact of increasing the number of workers on the training speed in Figure~\ref{fig:scale-speed}. We measure training speed by the number of training steps executed per second. %
Due to the proprietary nature of the data, we report training speed normalized by the average training speed for one worker. 
The training scales up linearly in the initial phase, however when the pool of workers becomes too large, we are faced with two confounding factors. First is the communication overhead for gradient updates. Second, more workers stay idle waiting for data to become available. Therefore, the I/O costs begin to dominate, and the total training time stagnates and even slows down, as seen in the $20+$ worker region of Figure~\ref{fig:scale-speed}.

\subsubsection{Effect of scaling on metrics}
\begin{figure}[t]
\centering
\includegraphics[width=0.8\linewidth,keepaspectratio]{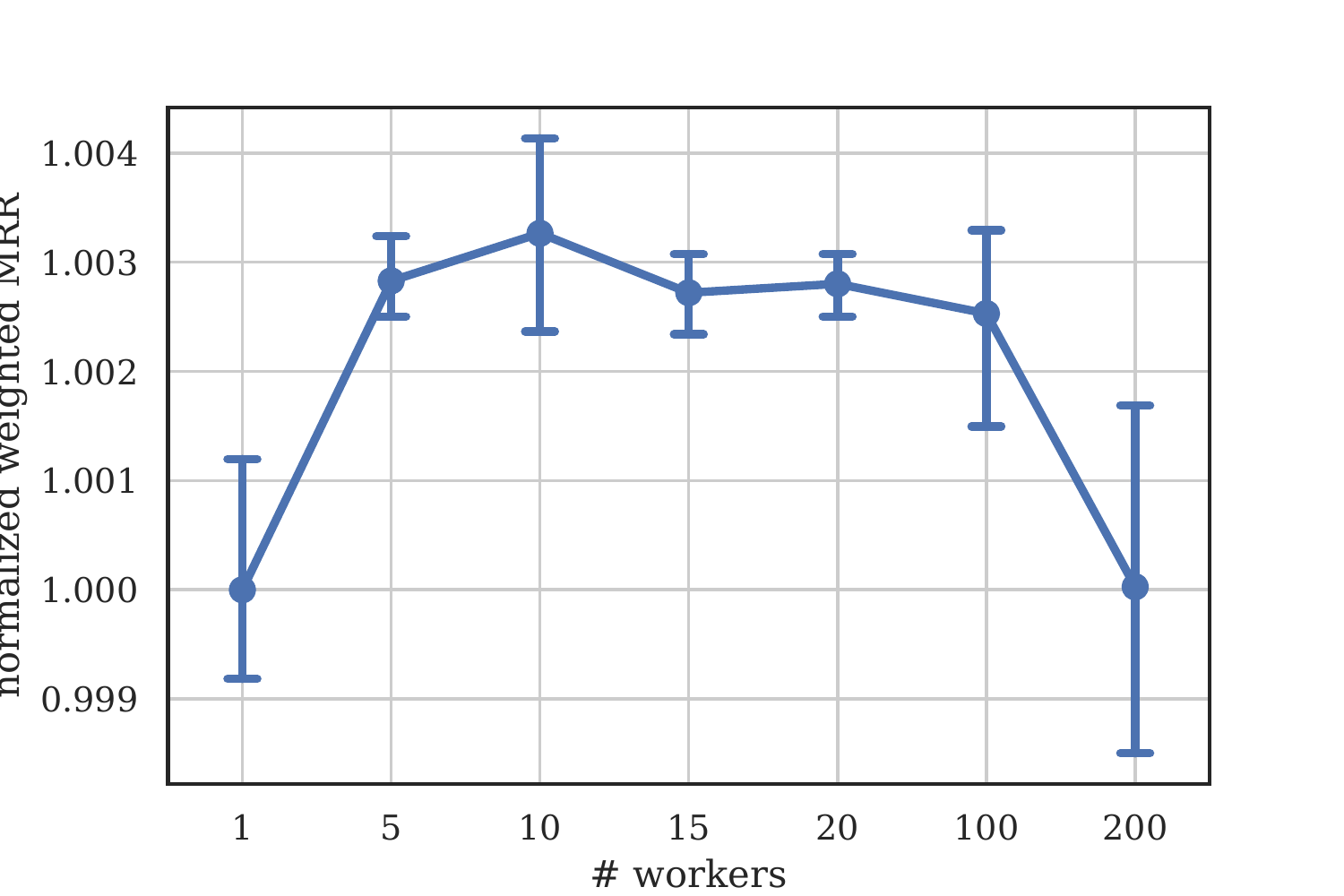}
\caption{Normalized weighted Mean Reciprocal Rank for Gmail Search as a function of number of workers.}
\label{fig:scale-accuracy}
\vspace{-8pt}
\end{figure}

In Figure~\ref{fig:scale-accuracy}, we examine the impact of increasing the number of workers on weighted MRR. Due to the proprietary nature of the data, we report the metric value normalized by the average metric for one worker. 

We observe that scaling does not generally have a significant impact on MRR, with two exceptions: using (a) a single worker, and (b) a large worker pool. We believe that the former requires more training cycles to achieve a comparable MRR. In the latter setting, gradient updates aggregated from too many workers become inaccurate, sending the model in a direction that is not well-aligned with the true gradient. However, in both cases, the overall effect on MRR is very small (roughly $0.03\%$), demonstrating the robustness of scaling with respect to model performance.

\section{Conclusion} \label{sec:conclusion}
In this paper we introduced TensorFlow Ranking---a scalable learning-to-rank library in TensorFlow. The library is highly configurable and has easy-to-use APIs for scoring mechanisms, loss functions, and evaluation metrics. 
Unlike the existing learning-to-rank open source packages which are designed for small datasets, TensorFlow Ranking can be used to solve real-world, large-scale ranking problems with hundreds of millions of training examples, and scales well to large clusters. TensorFlow Ranking is already deployed in several production systems at Google, and in this paper we empirically demonstrate its effectiveness for Gmail search and Quick Access in Google Drive~\cite{tata2017quick}. Our experiments show that TensorFlow Ranking can (a) leverage listwise loss functions, (b) effectively incorporate sparse features through embeddings, and (c) scale up without a significant drop in metrics.
TensorFlow Ranking is available to the open source community, and we hope that it facilitates further academic research and industrial applications.

\section{Acknowledgements}
  We thank the members of the TensorFlow team for their advice and support: Alexandre Passos, Mustafa Ispir, Karmel Allison, Martin Wicke, Clemens Mewald and others. We extend our special thanks to our collaborators, interns and early adopters: Suming Chen, Zhen Qin, Chirag Sethi, Maryam Karimzadehgan, Makoto Uchida, Yan Zhu, Qingyao Ai, Brandon Tran, Donald Metzler, Mike Colagrosso, Patrick McGregor and many others at Google who helped in evaluating and testing the early versions of TF-Ranking.

\balance
\bibliographystyle{ACM-Reference-Format}
\bibliography{references} 


\begin{thebibliography}{48}


\ifx \showCODEN    \undefined \def \showCODEN     #1{\unskip}     \fi
\ifx \showDOI      \undefined \def \showDOI       #1{#1}\fi
\ifx \showISBNx    \undefined \def \showISBNx     #1{\unskip}     \fi
\ifx \showISBNxiii \undefined \def \showISBNxiii  #1{\unskip}     \fi
\ifx \showISSN     \undefined \def \showISSN      #1{\unskip}     \fi
\ifx \showLCCN     \undefined \def \showLCCN      #1{\unskip}     \fi
\ifx \shownote     \undefined \def \shownote      #1{#1}          \fi
\ifx \showarticletitle \undefined \def \showarticletitle #1{#1}   \fi
\ifx \showURL      \undefined \def \showURL       {\relax}        \fi
\providecommand\bibfield[2]{#2}
\providecommand\bibinfo[2]{#2}
\providecommand\natexlab[1]{#1}
\providecommand\showeprint[2][]{arXiv:#2}

\bibitem[\protect\citeauthoryear{Abadi, Barham, Chen, Chen, Davis, Dean, Devin,
  Ghemawat, Irving, Isard, et~al\mbox{.}}{Abadi et~al\mbox{.}}{2016}]%
        {abadi2016tensorflow}
\bibfield{author}{\bibinfo{person}{Mart{\'\i}n Abadi}, \bibinfo{person}{Paul
  Barham}, \bibinfo{person}{Jianmin Chen}, \bibinfo{person}{Zhifeng Chen},
  \bibinfo{person}{Andy Davis}, \bibinfo{person}{Jeffrey Dean},
  \bibinfo{person}{Matthieu Devin}, \bibinfo{person}{Sanjay Ghemawat},
  \bibinfo{person}{Geoffrey Irving}, \bibinfo{person}{Michael Isard},
  {et~al\mbox{.}}} \bibinfo{year}{2016}\natexlab{}.
\newblock \showarticletitle{Tensorflow: a system for large-scale machine
  learning.}. In \bibinfo{booktitle}{\emph{12th {USENIX} Symposium on Operating
  Systems Design and Implementation}}. \bibinfo{pages}{265--283}.
\newblock


\bibitem[\protect\citeauthoryear{Ai, Mao, Liu, and Croft}{Ai
  et~al\mbox{.}}{2018a}]%
        {ai2018unbiased}
\bibfield{author}{\bibinfo{person}{Qingyao Ai}, \bibinfo{person}{Jiaxin Mao},
  \bibinfo{person}{Yiqun Liu}, {and} \bibinfo{person}{W~Bruce Croft}.}
  \bibinfo{year}{2018}\natexlab{a}.
\newblock \showarticletitle{Unbiased Learning to Rank: Theory and Practice}. In
  \bibinfo{booktitle}{\emph{2018 {ACM} {SIGIR} International Conference on
  Theory of Information Retrieval}}. \bibinfo{pages}{1--2}.
\newblock


\bibitem[\protect\citeauthoryear{Ai, Wang, Golbandi, Bendersky, and Najork}{Ai
  et~al\mbox{.}}{2018b}]%
        {ai2018groupwise}
\bibfield{author}{\bibinfo{person}{Qingyao Ai}, \bibinfo{person}{Xuanhui Wang},
  \bibinfo{person}{Nadav Golbandi}, \bibinfo{person}{Michael Bendersky}, {and}
  \bibinfo{person}{Marc Najork}.} \bibinfo{year}{2018}\natexlab{b}.
\newblock \showarticletitle{Learning Groupwise Scoring Functions Using Deep
  Neural Networks}.
\newblock \bibinfo{journal}{\emph{arXiv preprint arXiv:1811.04415}}
  (\bibinfo{year}{2018}).
\newblock


\bibitem[\protect\citeauthoryear{Bengio, Courville, and Vincent}{Bengio
  et~al\mbox{.}}{2013}]%
        {bengio2013representation}
\bibfield{author}{\bibinfo{person}{Yoshua Bengio}, \bibinfo{person}{Aaron
  Courville}, {and} \bibinfo{person}{Pascal Vincent}.}
  \bibinfo{year}{2013}\natexlab{}.
\newblock \showarticletitle{Representation learning: A review and new
  perspectives}.
\newblock \bibinfo{journal}{\emph{IEEE Transactions on Pattern Analysis and
  Machine Intelligence}} \bibinfo{volume}{35}, \bibinfo{number}{8}
  (\bibinfo{year}{2013}), \bibinfo{pages}{1798--1828}.
\newblock


\bibitem[\protect\citeauthoryear{Burges, Shaked, Renshaw, Lazier, Deeds,
  Hamilton, and Hullender}{Burges et~al\mbox{.}}{2005}]%
        {burges2005learning}
\bibfield{author}{\bibinfo{person}{Chris Burges}, \bibinfo{person}{Tal Shaked},
  \bibinfo{person}{Erin Renshaw}, \bibinfo{person}{Ari Lazier},
  \bibinfo{person}{Matt Deeds}, \bibinfo{person}{Nicole Hamilton}, {and}
  \bibinfo{person}{Greg Hullender}.} \bibinfo{year}{2005}\natexlab{}.
\newblock \showarticletitle{Learning to rank using gradient descent}. In
  \bibinfo{booktitle}{\emph{22nd International Conference on Machine
  Learning}}. \bibinfo{pages}{89--96}.
\newblock


\bibitem[\protect\citeauthoryear{Burges}{Burges}{2010}]%
        {burges2010ranknet}
\bibfield{author}{\bibinfo{person}{Christopher~J.C. Burges}.}
  \bibinfo{year}{2010}\natexlab{}.
\newblock \bibinfo{booktitle}{\emph{From {RankNet} to {LambdaRank} to
  {LambdaMART}: An Overview}}.
\newblock \bibinfo{type}{{T}echnical {R}eport} Technical Report MSR-TR-2010-82.
  \bibinfo{institution}{Microsoft Research}.
\newblock


\bibitem[\protect\citeauthoryear{Cao, Qin, Liu, Tsai, and Li}{Cao
  et~al\mbox{.}}{2007}]%
        {cao2007learning}
\bibfield{author}{\bibinfo{person}{Zhe Cao}, \bibinfo{person}{Tao Qin},
  \bibinfo{person}{Tie-Yan Liu}, \bibinfo{person}{Ming-Feng Tsai}, {and}
  \bibinfo{person}{Hang Li}.} \bibinfo{year}{2007}\natexlab{}.
\newblock \showarticletitle{Learning to rank: from pairwise approach to
  listwise approach}. In \bibinfo{booktitle}{\emph{24th International
  Conference on Machine Learning}}. \bibinfo{pages}{129--136}.
\newblock


\bibitem[\protect\citeauthoryear{Chapelle, Metzler, Zhang, and
  Grinspan}{Chapelle et~al\mbox{.}}{2009}]%
        {Chapelle+al:2009}
\bibfield{author}{\bibinfo{person}{Olivier Chapelle}, \bibinfo{person}{Donald
  Metzler}, \bibinfo{person}{Ya Zhang}, {and} \bibinfo{person}{Pierre
  Grinspan}.} \bibinfo{year}{2009}\natexlab{}.
\newblock \showarticletitle{Expected Reciprocal Rank for Graded Relevance}. In
  \bibinfo{booktitle}{\emph{18th {ACM} Conference on Information and Knowledge
  Management}}. \bibinfo{pages}{621--630}.
\newblock


\bibitem[\protect\citeauthoryear{Chauvin and Rumelhart}{Chauvin and
  Rumelhart}{2013}]%
        {chauvin2013backpropagation}
\bibfield{author}{\bibinfo{person}{Yves Chauvin} {and} \bibinfo{person}{David~E
  Rumelhart}.} \bibinfo{year}{2013}\natexlab{}.
\newblock \bibinfo{booktitle}{\emph{Backpropagation: theory, architectures, and
  applications}}.
\newblock \bibinfo{publisher}{Psychology Press}.
\newblock


\bibitem[\protect\citeauthoryear{Chen, Li, Li, Lin, Wang, Wang, Xiao, Xu,
  Zhang, and Zhang}{Chen et~al\mbox{.}}{2015}]%
        {chen2015mxnet}
\bibfield{author}{\bibinfo{person}{Tianqi Chen}, \bibinfo{person}{Mu Li},
  \bibinfo{person}{Yutian Li}, \bibinfo{person}{Min Lin},
  \bibinfo{person}{Naiyan Wang}, \bibinfo{person}{Minjie Wang},
  \bibinfo{person}{Tianjun Xiao}, \bibinfo{person}{Bing Xu},
  \bibinfo{person}{Chiyuan Zhang}, {and} \bibinfo{person}{Zheng Zhang}.}
  \bibinfo{year}{2015}\natexlab{}.
\newblock \showarticletitle{MXNet: {A} Flexible and Efficient Machine Learning
  Library for Heterogeneous Distributed Systems}.
\newblock \bibinfo{journal}{\emph{arXiv preprint arXiv:1512.01274}}
  (\bibinfo{year}{2015}).
\newblock


\bibitem[\protect\citeauthoryear{Chen, Liu, Lan, Ma, and Li}{Chen
  et~al\mbox{.}}{2009}]%
        {NIPS2009_3708}
\bibfield{author}{\bibinfo{person}{Wei Chen}, \bibinfo{person}{Tie-Yan Liu},
  \bibinfo{person}{Yanyan Lan}, \bibinfo{person}{Zhi-Ming Ma}, {and}
  \bibinfo{person}{Hang Li}.} \bibinfo{year}{2009}\natexlab{}.
\newblock \showarticletitle{Ranking Measures and Loss Functions in Learning to
  Rank}.
\newblock In \bibinfo{booktitle}{\emph{Advances in Neural Information
  Processing Systems}}. \bibinfo{pages}{315--323}.
\newblock


\bibitem[\protect\citeauthoryear{Cheng, Haque, Hong, Ispir, Mewald, Polosukhin,
  Roumpos, Sculley, Smith, Soergel, et~al\mbox{.}}{Cheng et~al\mbox{.}}{2017}]%
        {cheng2017tensorflow}
\bibfield{author}{\bibinfo{person}{Heng-Tze Cheng}, \bibinfo{person}{Zakaria
  Haque}, \bibinfo{person}{Lichan Hong}, \bibinfo{person}{Mustafa Ispir},
  \bibinfo{person}{Clemens Mewald}, \bibinfo{person}{Illia Polosukhin},
  \bibinfo{person}{Georgios Roumpos}, \bibinfo{person}{D Sculley},
  \bibinfo{person}{Jamie Smith}, \bibinfo{person}{David Soergel},
  {et~al\mbox{.}}} \bibinfo{year}{2017}\natexlab{}.
\newblock \showarticletitle{Tensorflow estimators: Managing simplicity vs.
  flexibility in high-level machine learning frameworks}. In
  \bibinfo{booktitle}{\emph{23rd {ACM} {SIGKDD} International Conference on
  Knowledge Discovery and Data Mining}}. \bibinfo{pages}{1763--1771}.
\newblock


\bibitem[\protect\citeauthoryear{Chu and Ghahramani}{Chu and
  Ghahramani}{2005}]%
        {Chu:2005:PLG:1102351.1102369}
\bibfield{author}{\bibinfo{person}{Wei Chu} {and} \bibinfo{person}{Zoubin
  Ghahramani}.} \bibinfo{year}{2005}\natexlab{}.
\newblock \showarticletitle{Preference Learning with Gaussian Processes}. In
  \bibinfo{booktitle}{\emph{22nd International Conference on Machine
  Learning}}. \bibinfo{pages}{137--144}.
\newblock


\bibitem[\protect\citeauthoryear{Craswell}{Craswell}{2009}]%
        {craswell2009mean}
\bibfield{author}{\bibinfo{person}{Nick Craswell}.}
  \bibinfo{year}{2009}\natexlab{}.
\newblock \showarticletitle{Mean reciprocal rank}.
\newblock In \bibinfo{booktitle}{\emph{Encyclopedia of Database Systems}}.
  \bibinfo{publisher}{Springer}, \bibinfo{pages}{1703--1703}.
\newblock


\bibitem[\protect\citeauthoryear{Duchi, Hazan, and Singer}{Duchi
  et~al\mbox{.}}{2011}]%
        {duchi2011adaptive}
\bibfield{author}{\bibinfo{person}{John Duchi}, \bibinfo{person}{Elad Hazan},
  {and} \bibinfo{person}{Yoram Singer}.} \bibinfo{year}{2011}\natexlab{}.
\newblock \showarticletitle{Adaptive subgradient methods for online learning
  and stochastic optimization}.
\newblock \bibinfo{journal}{\emph{Journal of Machine Learning Research}}
  \bibinfo{volume}{12} (\bibinfo{date}{July} \bibinfo{year}{2011}),
  \bibinfo{pages}{2121--2159}.
\newblock


\bibitem[\protect\citeauthoryear{Friedman}{Friedman}{2001}]%
        {friedman2001greedy}
\bibfield{author}{\bibinfo{person}{Jerome~H Friedman}.}
  \bibinfo{year}{2001}\natexlab{}.
\newblock \showarticletitle{Greedy function approximation: a gradient boosting
  machine}.
\newblock \bibinfo{journal}{\emph{Annals of Statistics}} \bibinfo{volume}{29},
  \bibinfo{number}{5} (\bibinfo{year}{2001}), \bibinfo{pages}{1189--1232}.
\newblock


\bibitem[\protect\citeauthoryear{Fuhr}{Fuhr}{1989}]%
        {Fuhr:1989:OPR:65943.65944}
\bibfield{author}{\bibinfo{person}{Norbert Fuhr}.}
  \bibinfo{year}{1989}\natexlab{}.
\newblock \showarticletitle{Optimum Polynomial Retrieval Functions Based on the
  Probability Ranking Principle}.
\newblock \bibinfo{journal}{\emph{ACM Transactions on Information Systems}}
  \bibinfo{volume}{7}, \bibinfo{number}{3} (\bibinfo{year}{1989}),
  \bibinfo{pages}{183--204}.
\newblock


\bibitem[\protect\citeauthoryear{Gey}{Gey}{1994}]%
        {Gey:1994:IPR:188490.188560}
\bibfield{author}{\bibinfo{person}{Fredric~C. Gey}.}
  \bibinfo{year}{1994}\natexlab{}.
\newblock \showarticletitle{Inferring Probability of Relevance Using the Method
  of Logistic Regression}. In \bibinfo{booktitle}{\emph{17th Annual
  International ACM SIGIR Conference on Research and Development in Information
  Retrieval}}. \bibinfo{pages}{222--231}.
\newblock


\bibitem[\protect\citeauthoryear{Goodfellow, Bengio, Courville, and
  Bengio}{Goodfellow et~al\mbox{.}}{2016}]%
        {goodfellow2016deep}
\bibfield{author}{\bibinfo{person}{Ian Goodfellow}, \bibinfo{person}{Yoshua
  Bengio}, \bibinfo{person}{Aaron Courville}, {and} \bibinfo{person}{Yoshua
  Bengio}.} \bibinfo{year}{2016}\natexlab{}.
\newblock \bibinfo{booktitle}{\emph{Deep Learning}}.
\newblock \bibinfo{publisher}{MIT Press Cambridge}.
\newblock


\bibitem[\protect\citeauthoryear{J{\"a}rvelin and
  Kek{\"a}l{\"a}inen}{J{\"a}rvelin and Kek{\"a}l{\"a}inen}{2002}]%
        {jarvelin2002cumulated}
\bibfield{author}{\bibinfo{person}{Kalervo J{\"a}rvelin} {and}
  \bibinfo{person}{Jaana Kek{\"a}l{\"a}inen}.} \bibinfo{year}{2002}\natexlab{}.
\newblock \showarticletitle{Cumulated gain-based evaluation of {IR}
  techniques}.
\newblock \bibinfo{journal}{\emph{ACM Transactions on Information Systems}}
  \bibinfo{volume}{20}, \bibinfo{number}{4} (\bibinfo{year}{2002}),
  \bibinfo{pages}{422--446}.
\newblock


\bibitem[\protect\citeauthoryear{Jia, Shelhamer, Donahue, Karayev, Long,
  Girshick, Guadarrama, and Darrell}{Jia et~al\mbox{.}}{2014}]%
        {jia2014caffe}
\bibfield{author}{\bibinfo{person}{Yangqing Jia}, \bibinfo{person}{Evan
  Shelhamer}, \bibinfo{person}{Jeff Donahue}, \bibinfo{person}{Sergey Karayev},
  \bibinfo{person}{Jonathan Long}, \bibinfo{person}{Ross Girshick},
  \bibinfo{person}{Sergio Guadarrama}, {and} \bibinfo{person}{Trevor Darrell}.}
  \bibinfo{year}{2014}\natexlab{}.
\newblock \showarticletitle{Caffe: Convolutional architecture for fast feature
  embedding}. In \bibinfo{booktitle}{\emph{22nd ACM International Conference on
  Multimedia}}. \bibinfo{pages}{675--678}.
\newblock


\bibitem[\protect\citeauthoryear{Joachims}{Joachims}{2002}]%
        {Joachims:2002}
\bibfield{author}{\bibinfo{person}{Thorsten Joachims}.}
  \bibinfo{year}{2002}\natexlab{}.
\newblock \showarticletitle{Optimizing Search Engines Using Clickthrough Data}.
  In \bibinfo{booktitle}{\emph{8th ACM SIGKDD International Conference on
  Knowledge Discovery and Data Mining}}. \bibinfo{pages}{133--142}.
\newblock


\bibitem[\protect\citeauthoryear{Joachims}{Joachims}{2006}]%
        {joachims2006training}
\bibfield{author}{\bibinfo{person}{Thorsten Joachims}.}
  \bibinfo{year}{2006}\natexlab{}.
\newblock \showarticletitle{Training linear SVMs in linear time}. In
  \bibinfo{booktitle}{\emph{12th ACM SIGKDD International Conference on
  Knowledge Discovery and Data Mining}}. \bibinfo{pages}{217--226}.
\newblock


\bibitem[\protect\citeauthoryear{Joachims, Granka, Pan, Hembrooke, and
  Gay}{Joachims et~al\mbox{.}}{2005}]%
        {Joachims+al:2005}
\bibfield{author}{\bibinfo{person}{Thorsten Joachims}, \bibinfo{person}{Laura
  Granka}, \bibinfo{person}{Bing Pan}, \bibinfo{person}{Helene Hembrooke},
  {and} \bibinfo{person}{Geri Gay}.} \bibinfo{year}{2005}\natexlab{}.
\newblock \showarticletitle{Accurately Interpreting Clickthrough Data As
  Implicit Feedback}. In \bibinfo{booktitle}{\emph{28th Annual International
  ACM SIGIR Conference on Research and Development in Information Retrieval}}.
  \bibinfo{pages}{154--161}.
\newblock


\bibitem[\protect\citeauthoryear{Joachims, Swaminathan, and Schnabel}{Joachims
  et~al\mbox{.}}{2017}]%
        {Joachims:WSDM17}
\bibfield{author}{\bibinfo{person}{Thorsten Joachims}, \bibinfo{person}{Adith
  Swaminathan}, {and} \bibinfo{person}{Tobias Schnabel}.}
  \bibinfo{year}{2017}\natexlab{}.
\newblock \showarticletitle{Unbiased Learning-to-Rank with Biased Feedback}. In
  \bibinfo{booktitle}{\emph{10th ACM International Conference on Web Search and
  Data Mining}}. \bibinfo{pages}{781--789}.
\newblock


\bibitem[\protect\citeauthoryear{Ke, Meng, Finley, Wang, Chen, Ma, Ye, and
  Liu}{Ke et~al\mbox{.}}{2017}]%
        {ke2017lightgbm}
\bibfield{author}{\bibinfo{person}{Guolin Ke}, \bibinfo{person}{Qi Meng},
  \bibinfo{person}{Thomas Finley}, \bibinfo{person}{Taifeng Wang},
  \bibinfo{person}{Wei Chen}, \bibinfo{person}{Weidong Ma},
  \bibinfo{person}{Qiwei Ye}, {and} \bibinfo{person}{Tie-Yan Liu}.}
  \bibinfo{year}{2017}\natexlab{}.
\newblock \showarticletitle{{LightGBM}: A highly efficient gradient boosting
  decision tree}. In \bibinfo{booktitle}{\emph{Advances in Neural Information
  Processing Systems 30}}. \bibinfo{pages}{3146--3154}.
\newblock


\bibitem[\protect\citeauthoryear{LeCun and Bengio}{LeCun and Bengio}{1995}]%
        {lecun1995convolutional}
\bibfield{author}{\bibinfo{person}{Yann LeCun} {and} \bibinfo{person}{Yoshua
  Bengio}.} \bibinfo{year}{1995}\natexlab{}.
\newblock \showarticletitle{Convolutional networks for images, speech, and time
  series}.
\newblock In \bibinfo{booktitle}{\emph{The Handbook of Brain Theory and Neural
  Networks}}, \bibfield{editor}{\bibinfo{person}{Michael~A Arbib}} (Ed.).
  \bibinfo{publisher}{MIT Press}, \bibinfo{pages}{255--258}.
\newblock


\bibitem[\protect\citeauthoryear{Li}{Li}{2011}]%
        {li2011learning}
\bibfield{author}{\bibinfo{person}{Hang Li}.} \bibinfo{year}{2011}\natexlab{}.
\newblock \showarticletitle{Learning to rank for information retrieval and
  natural language processing}.
\newblock \bibinfo{journal}{\emph{Synthesis Lectures on Human Language
  Technologies}} \bibinfo{volume}{4}, \bibinfo{number}{1}
  (\bibinfo{year}{2011}), \bibinfo{pages}{1--113}.
\newblock


\bibitem[\protect\citeauthoryear{Metzler, Croft, and Mccallum}{Metzler
  et~al\mbox{.}}{2005}]%
        {metzler2005directMaximization}
\bibfield{author}{\bibinfo{person}{Donald~A Metzler}, \bibinfo{person}{W~Bruce
  Croft}, {and} \bibinfo{person}{Andrew Mccallum}.}
  \bibinfo{year}{2005}\natexlab{}.
\newblock \bibinfo{booktitle}{\emph{Direct maximization of rank-based metrics
  for information retrieval}}.
\newblock \bibinfo{type}{CIIR report} 429. \bibinfo{institution}{University of
  Massachusetts}.
\newblock


\bibitem[\protect\citeauthoryear{Mikolov, Sutskever, Chen, Corrado, and
  Dean}{Mikolov et~al\mbox{.}}{2013}]%
        {mikolov2013word2vec}
\bibfield{author}{\bibinfo{person}{Tomas Mikolov}, \bibinfo{person}{Ilya
  Sutskever}, \bibinfo{person}{Kai Chen}, \bibinfo{person}{Greg~S Corrado},
  {and} \bibinfo{person}{Jeff Dean}.} \bibinfo{year}{2013}\natexlab{}.
\newblock \showarticletitle{Distributed Representations of Words and Phrases
  and their Compositionality}.
\newblock In \bibinfo{booktitle}{\emph{Advances in Neural Information
  Processing Systems 26}}. \bibinfo{pages}{3111--3119}.
\newblock


\bibitem[\protect\citeauthoryear{Mitra and Craswell}{Mitra and
  Craswell}{2017}]%
        {mitra2017neural}
\bibfield{author}{\bibinfo{person}{Bhaskar Mitra} {and} \bibinfo{person}{Nick
  Craswell}.} \bibinfo{year}{2017}\natexlab{}.
\newblock \showarticletitle{Neural Models for Information Retrieval}.
\newblock \bibinfo{journal}{\emph{arXiv preprint arXiv:1705.01509}}
  (\bibinfo{year}{2017}).
\newblock


\bibitem[\protect\citeauthoryear{Nair and Hinton}{Nair and Hinton}{2010}]%
        {nair2010rectified}
\bibfield{author}{\bibinfo{person}{Vinod Nair} {and}
  \bibinfo{person}{Geoffrey~E Hinton}.} \bibinfo{year}{2010}\natexlab{}.
\newblock \showarticletitle{Rectified linear units improve restricted
  {Boltzmann} machines}. In \bibinfo{booktitle}{\emph{27th International
  Conference on Machine Learning}}. \bibinfo{pages}{807--814}.
\newblock


\bibitem[\protect\citeauthoryear{Olston, Fiedel, Gorovoy, Harmsen, Lao, Li,
  Rajashekhar, Ramesh, and Soyke}{Olston et~al\mbox{.}}{2017}]%
        {olston2017tensorflow}
\bibfield{author}{\bibinfo{person}{Christopher Olston}, \bibinfo{person}{Noah
  Fiedel}, \bibinfo{person}{Kiril Gorovoy}, \bibinfo{person}{Jeremiah Harmsen},
  \bibinfo{person}{Li Lao}, \bibinfo{person}{Fangwei Li}, \bibinfo{person}{Vinu
  Rajashekhar}, \bibinfo{person}{Sukriti Ramesh}, {and} \bibinfo{person}{Jordan
  Soyke}.} \bibinfo{year}{2017}\natexlab{}.
\newblock \showarticletitle{TensorFlow-Serving: Flexible, high-performance ML
  serving}.
\newblock \bibinfo{journal}{\emph{arXiv preprint arXiv:1712.06139}}
  (\bibinfo{year}{2017}).
\newblock


\bibitem[\protect\citeauthoryear{Paszke, Gross, Chintala, Chanan, Yang, DeVito,
  Lin, Desmaison, Antiga, and Lerer}{Paszke et~al\mbox{.}}{2017}]%
        {paszke2017automatic}
\bibfield{author}{\bibinfo{person}{Adam Paszke}, \bibinfo{person}{Sam Gross},
  \bibinfo{person}{Soumith Chintala}, \bibinfo{person}{Gregory Chanan},
  \bibinfo{person}{Edward Yang}, \bibinfo{person}{Zachary DeVito},
  \bibinfo{person}{Zeming Lin}, \bibinfo{person}{Alban Desmaison},
  \bibinfo{person}{Luca Antiga}, {and} \bibinfo{person}{Adam Lerer}.}
  \bibinfo{year}{2017}\natexlab{}.
\newblock \showarticletitle{Automatic differentiation in PyTorch}. In
  \bibinfo{booktitle}{\emph{AutoDiff Workshop at NIPS 2017}}.
\newblock


\bibitem[\protect\citeauthoryear{Qin, Liu, and Li}{Qin et~al\mbox{.}}{2010}]%
        {Qin:2010:GAF:1842549.1842572}
\bibfield{author}{\bibinfo{person}{Tao Qin}, \bibinfo{person}{Tie-Yan Liu},
  {and} \bibinfo{person}{Hang Li}.} \bibinfo{year}{2010}\natexlab{}.
\newblock \showarticletitle{A General Approximation Framework for Direct
  Optimization of Information Retrieval Measures}.
\newblock \bibinfo{journal}{\emph{Information Retrieval}} \bibinfo{volume}{13},
  \bibinfo{number}{4} (\bibinfo{year}{2010}), \bibinfo{pages}{375--397}.
\newblock


\bibitem[\protect\citeauthoryear{Silfverberg, Mao, and Hulden}{Silfverberg
  et~al\mbox{.}}{2018}]%
        {silfverberg2018sound}
\bibfield{author}{\bibinfo{person}{Miikka~P Silfverberg},
  \bibinfo{person}{Lingshuang~Jack Mao}, {and} \bibinfo{person}{Mans Hulden}.}
  \bibinfo{year}{2018}\natexlab{}.
\newblock \showarticletitle{Sound Analogies with Phoneme Embeddings}.
\newblock \bibinfo{journal}{\emph{Proc. of the Society for Computation in
  Linguistics (SCiL)}} (\bibinfo{year}{2018}), \bibinfo{pages}{136--144}.
\newblock


\bibitem[\protect\citeauthoryear{Srivastava, Hinton, Krizhevsky, Sutskever, and
  Salakhutdinov}{Srivastava et~al\mbox{.}}{2014}]%
        {srivastava2014dropout}
\bibfield{author}{\bibinfo{person}{Nitish Srivastava},
  \bibinfo{person}{Geoffrey Hinton}, \bibinfo{person}{Alex Krizhevsky},
  \bibinfo{person}{Ilya Sutskever}, {and} \bibinfo{person}{Ruslan
  Salakhutdinov}.} \bibinfo{year}{2014}\natexlab{}.
\newblock \showarticletitle{Dropout: a simple way to prevent neural networks
  from overfitting}.
\newblock \bibinfo{journal}{\emph{Journal of Machine Learning Research}}
  \bibinfo{volume}{15}, \bibinfo{number}{1} (\bibinfo{year}{2014}),
  \bibinfo{pages}{1929--1958}.
\newblock


\bibitem[\protect\citeauthoryear{Tata, Popescul, Najork, Colagrosso, Gibbons,
  Green, Mah, Smith, Garg, Meyer, et~al\mbox{.}}{Tata et~al\mbox{.}}{2017}]%
        {tata2017quick}
\bibfield{author}{\bibinfo{person}{Sandeep Tata}, \bibinfo{person}{Alexandrin
  Popescul}, \bibinfo{person}{Marc Najork}, \bibinfo{person}{Mike Colagrosso},
  \bibinfo{person}{Julian Gibbons}, \bibinfo{person}{Alan Green},
  \bibinfo{person}{Alexandre Mah}, \bibinfo{person}{Michael Smith},
  \bibinfo{person}{Divanshu Garg}, \bibinfo{person}{Cayden Meyer},
  {et~al\mbox{.}}} \bibinfo{year}{2017}\natexlab{}.
\newblock \showarticletitle{Quick {Access}: Building a Smart Experience for
  {Google} {Drive}}. In \bibinfo{booktitle}{\emph{23rd ACM SIGKDD International
  Conference on Knowledge Discovery and Data Mining}}.
  \bibinfo{pages}{1643--1651}.
\newblock


\bibitem[\protect\citeauthoryear{Taylor, Guiver, Robertson, and Minka}{Taylor
  et~al\mbox{.}}{2008}]%
        {Taylor+al:2008}
\bibfield{author}{\bibinfo{person}{Michael Taylor}, \bibinfo{person}{John
  Guiver}, \bibinfo{person}{Stephen Robertson}, {and} \bibinfo{person}{Tom
  Minka}.} \bibinfo{year}{2008}\natexlab{}.
\newblock \showarticletitle{SoftRank: Optimizing Non-smooth Rank Metrics}. In
  \bibinfo{booktitle}{\emph{1st International Conference on Web Search and Web
  Data Mining}}. \bibinfo{pages}{77--86}.
\newblock


\bibitem[\protect\citeauthoryear{Wang, Bendersky, Metzler, and Najork}{Wang
  et~al\mbox{.}}{2016}]%
        {Wang+al:2016}
\bibfield{author}{\bibinfo{person}{Xuanhui Wang}, \bibinfo{person}{Michael
  Bendersky}, \bibinfo{person}{Donald Metzler}, {and} \bibinfo{person}{Marc
  Najork}.} \bibinfo{year}{2016}\natexlab{}.
\newblock \showarticletitle{Learning to Rank with Selection Bias in Personal
  Search}. In \bibinfo{booktitle}{\emph{39th International ACM SIGIR conference
  on Research and Development in Information Retrieval}}.
  \bibinfo{pages}{115--124}.
\newblock


\bibitem[\protect\citeauthoryear{Wang, Golbandi, Bendersky, Metzler, and
  Najork}{Wang et~al\mbox{.}}{2018a}]%
        {Wang+al:2018}
\bibfield{author}{\bibinfo{person}{Xuanhui Wang}, \bibinfo{person}{Nadav
  Golbandi}, \bibinfo{person}{Michael Bendersky}, \bibinfo{person}{Donald
  Metzler}, {and} \bibinfo{person}{Marc Najork}.}
  \bibinfo{year}{2018}\natexlab{a}.
\newblock \showarticletitle{Position Bias Estimation for Unbiased Learning to
  Rank in Personal Search}. In \bibinfo{booktitle}{\emph{11th ACM International
  Conference on Web Search and Data Mining}}. \bibinfo{pages}{610 --618}.
\newblock


\bibitem[\protect\citeauthoryear{Wang, Li, Golbandi, Bendersky, and
  Najork}{Wang et~al\mbox{.}}{2018b}]%
        {wang2018lambdaloss}
\bibfield{author}{\bibinfo{person}{Xuanhui Wang}, \bibinfo{person}{Cheng Li},
  \bibinfo{person}{Nadav Golbandi}, \bibinfo{person}{Michael Bendersky}, {and}
  \bibinfo{person}{Marc Najork}.} \bibinfo{year}{2018}\natexlab{b}.
\newblock \showarticletitle{The {LambdaLoss} Framework for Ranking Metric
  Optimization}. In \bibinfo{booktitle}{\emph{27th ACM International Conference
  on Information and Knowledge Management}}. \bibinfo{pages}{1313--1322}.
\newblock


\bibitem[\protect\citeauthoryear{Xia, Liu, Wang, Zhang, and Li}{Xia
  et~al\mbox{.}}{2008}]%
        {xia2008listmle}
\bibfield{author}{\bibinfo{person}{Fen Xia}, \bibinfo{person}{Tie-Yan Liu},
  \bibinfo{person}{Jue Wang}, \bibinfo{person}{Wensheng Zhang}, {and}
  \bibinfo{person}{Hang Li}.} \bibinfo{year}{2008}\natexlab{}.
\newblock \showarticletitle{Listwise Approach to Learning to Rank: Theory and
  Algorithm}. In \bibinfo{booktitle}{\emph{25th International Conference on
  Machine Learning}}. \bibinfo{pages}{1192--1199}.
\newblock


\bibitem[\protect\citeauthoryear{Xiong, Dai, Callan, Liu, and Power}{Xiong
  et~al\mbox{.}}{2017}]%
        {Xiong+al:2017}
\bibfield{author}{\bibinfo{person}{Chenyan Xiong}, \bibinfo{person}{Zhuyun
  Dai}, \bibinfo{person}{Jamie Callan}, \bibinfo{person}{Zhiyuan Liu}, {and}
  \bibinfo{person}{Russell Power}.} \bibinfo{year}{2017}\natexlab{}.
\newblock \showarticletitle{End-to-End Neural Ad-hoc Ranking with Kernel
  Pooling}. In \bibinfo{booktitle}{\emph{40th International ACM SIGIR
  Conference on Research and Development in Information Retrieval}}.
  \bibinfo{pages}{55--64}.
\newblock


\bibitem[\protect\citeauthoryear{Xu and Li}{Xu and Li}{2007}]%
        {Jun+Hang:2007}
\bibfield{author}{\bibinfo{person}{Jun Xu} {and} \bibinfo{person}{Hang Li}.}
  \bibinfo{year}{2007}\natexlab{}.
\newblock \showarticletitle{AdaRank: A Boosting Algorithm for Information
  Retrieval}. In \bibinfo{booktitle}{\emph{30th Annual International ACM SIGIR
  Conference on Research and Development in Information Retrieval}}.
  \bibinfo{pages}{391--398}.
\newblock


\bibitem[\protect\citeauthoryear{Yue, Patel, and Roehrig}{Yue
  et~al\mbox{.}}{2010}]%
        {Yue:2010:BPB:1772690.1772793}
\bibfield{author}{\bibinfo{person}{Yisong Yue}, \bibinfo{person}{Rajan Patel},
  {and} \bibinfo{person}{Hein Roehrig}.} \bibinfo{year}{2010}\natexlab{}.
\newblock \showarticletitle{Beyond Position Bias: Examining Result
  Attractiveness As a Source of Presentation Bias in Clickthrough Data}. In
  \bibinfo{booktitle}{\emph{19th International Conference on World Wide Web}}.
  \bibinfo{pages}{1011--1018}.
\newblock


\bibitem[\protect\citeauthoryear{Zamani, Bendersky, Wang, and Zhang}{Zamani
  et~al\mbox{.}}{2017}]%
        {Zamani+al:2017}
\bibfield{author}{\bibinfo{person}{Hamed Zamani}, \bibinfo{person}{Michael
  Bendersky}, \bibinfo{person}{Xuanhui Wang}, {and} \bibinfo{person}{Mingyang
  Zhang}.} \bibinfo{year}{2017}\natexlab{}.
\newblock \showarticletitle{Situational Context for Ranking in Personal
  Search}. In \bibinfo{booktitle}{\emph{26th International Conference on World
  Wide Web}}. \bibinfo{pages}{1531--1540}.
\newblock


\bibitem[\protect\citeauthoryear{Zhu}{Zhu}{2004}]%
        {zhu2004recall}
\bibfield{author}{\bibinfo{person}{Mu Zhu}.} \bibinfo{year}{2004}\natexlab{}.
\newblock \bibinfo{booktitle}{\emph{Recall, precision and average precision}}.
\newblock \bibinfo{type}{{T}echnical {R}eport}.
  \bibinfo{institution}{Department of Statistics and Actuarial Science,
  University of Waterloo}.
\newblock


\end{thebibliography}
\end{document}